\documentstyle[12pt]{article}
\global\parskip 6pt

\normalsize

\let\oldtheequation=\theequation
\def\doteqs#1{\setcounter{equation}{0}
            \def\theequation{{#1}.\oldtheequation}}

\newcounter{sxn}
\def\sx#1{\addtocounter{sxn}{1} \bigskip\medskip \goodbreak
\noindent{\large\bf
\centerline{\thesxn.~~#1}} \nobreak \medskip}
\def\sxn#1{\sx{#1} \doteqs{\thesxn}}

\newcounter{axn}

\def\br{\begin{thebibliography}{abc}}
\def\er{\end{thebibliography}}

\def\be{\begin{equation}}
\def\ee{\end{equation}}
\def\bea{\begin{eqnarray}}
\def\eea{\end{eqnarray}}
\newcommand{\bm}[1]{\mbox{\boldmath$#1$}}
\newcommand{\grad}{\bm \nabla}

\begin{document}
\begin{flushright}
\hfill{SINP-TNP/02-20}\\
\end{flushright}
\vspace*{1cm}
\thispagestyle{empty}
\begin{center}
{\large\bf Novel Quantum States of the Rational Calogero Models\\
Without the Confining Interaction}\\
\end{center}
\bigskip
\begin{center}
B. Basu-Mallick\footnote{Email: biru@theory.saha.ernet.in},
Pijush K. Ghosh\footnote{Email: pijush@theory.saha.ernet.in}
and Kumar S. Gupta\footnote{Email: gupta@theory.saha.ernet.in}\\
\vspace*{.75cm}
{\em Theory Group\\
Saha Institute of Nuclear Physics\\
1/AF Bidhannagar, Calcutta - 700064, India}\\
\end{center}
\vskip.5cm

\begin{abstract}
We show that the $N$-particle $A_{N-1}$ and $B_N$ rational Calogero models
without the harmonic interaction admit a new class of bound and scattering 
states. These states
owe their existence to the self-adjoint extensions of the corresponding 
Hamiltonians, labelled by  $e^{iz}$ where $z \in R $ (mod $2 \pi$).
It is shown that the new states appear for all values of $N$ and
for specific ranges of the coupling constants. Moreover, they
are shown to exist even in the excited sectors of the Calogero models.
The self-adjoint extension generically breaks the 
classical scaling symmetry, leading to quantum mechanical scaling anomaly.
The scaling symmetry can  
however be restored for certain values of the parameter $z$. We also 
generalize these results for many particle systems with classically scale
invariant long range interactions in arbitrary dimensions.

\end{abstract}

\vspace*{1.0cm}

\vspace*{2.5cm}

\newpage

\baselineskip 15pt

\sxn{Introduction} 

The spectrum of the $N$-particle one dimensional $A_{N-1}$
rational Calogero model without the
confining term has traditionally been described in terms of scattering
states
alone \cite{calo1}. It has however recently been shown \cite{kumar1} that
the spectrum of such a system admits negative energy bound states for
certain values of the system parameters. These bound states owe their
existence to the self-adjoint extension \cite{reed,hut} of the ``radial'' part
of the Hamiltonian. These self-adjoint extensions are labelled by 
$e^{iz}$ where $z \in R$. The parameter $z$ (mod $2 \pi$)  
classifies all possible boundary conditions for the
system. Indeed, the corresponding result for the two-body case has  
been known for a long time \cite{meetz}, which has recently found 
applications in the context of black holes \cite{trg,kumar2,moretti}.

In the analysis presented in Ref. \cite{kumar1}, the new bound states 
of the $A_{N-1}$ rational Calogero model were 
found to exist only for $N=3$ and $4$ and for the zero 
``angular momentum'' sector of the problem. This happened due to the fact
that the  wavefunction was required to vanish when any two particles
coincide. Such a boundary condition allows to smoothly extend 
the wavefunction for a given particle ordering 
to the whole of the N-particle configuration space corresponding to all
possible orderings. In this paper we observe that this boundary condition on
the wavefunction can be relaxed while still keeping it square integrable
for any given particle ordering. 
The wavefunction obtained in this way can then be extended to all other
particle orderings by using the permutation symmetry, although 
not in a smooth fashion. As a consequence of the more general boundary
condition, we are now able to find the new bound states for arbitrary values 
of particle number $N$. Even more remarkably, the self-adjoint extensions are  
now found to exist even in the higher ``angular momentum'' sectors of the
problem.

The scattering state solutions of the $A_{N-1}$ model in absence of the
confining term was already found by Calogero in Ref. \cite{calo1}. The range
of the coupling there was such that the corresponding Hamiltonian
$H_{A_{N-1}}$ was essentially self-adjoint. In this paper we have been able to 
extend that analysis for the parameter range where $H_{A_{N-1}}$ is not 
essentially self-adjoint but admits self-adjoint extensions. 
In particular, we have found a new class of   
scattering states for each choice of the self-adjoint parameter $z$.  
The phase shifts for these scattering states depend explicitly on choice of
$z$. Moreover, for generic values of $z$, the phase shift is found to depend
on the momentum as well.

The $A_{N-1}$ Calogero model without the confining term belongs to a class
of $N$-particle systems with long range interactions which
are classically scale invariant [9-15]. The scale invariance manifests itself 
through the absence of dimensionfull coupling constants. It is thus natural 
to ask if an analysis similar to that in Ref. \cite{kumar1} applies to this 
wider class of systems, including the higher dimensional cases. In this 
paper we address this issue and show that these systems indeed admit a new
class of bound and scattering states. These states appear due to the
self-adjoint extensions of the corresponding Hamiltonians.
As examples of this general formulation, we have analyzed the bound and
scattering state sectors of the $A_{N-1}$ and
$B_N$ Calogero models in one dimension. The $D_N$ and $C_N$ Calogero models 
in one dimension arise as special cases of the $B_N$ model.
In all these cases we have found that the new quantum states exist
for arbitrary values of $N$ and for suitable ranges of the coupling 
constants. Moreover, these states appear in the excited sectors as well.
The Calogero-Marchioro model \cite{cm} in $D$ dimensions has also been
analyzed to illustrate our formalism in dimensions higher than one. It has
been shown that the new states exist in this model as well with suitable 
restrictions on the system parameters.

The classically scale invariant systems considered here are such that 
$O(2,1)$ appears as their spectrum generating 
algebra \cite{jackiw1,fubini}. It is however known that the classical scaling 
symmetry may be broken in the presence of the self-adjoint extensions 
\cite{dhowker,jackiw}. We shall show  
that for the general class of systems considered here, the self-adjoint
extensions typically break the scale invariance at the quantum
level. This is evident by the appearance of 
the momentum dependence of phase shifts in the scattering sector and by the
appearance of bound states in certain cases.
The underlying reason behind this is that the domain of self-adjointness of
the associated Hamiltonian is not kept invariant by the scaling operator
\cite{jackiw,esteve,dipti,camblong}. Scale invariance
at the quantum level is recovered only for special choices of the parameter 
$z$. The systems considered here thus provide further examples where scaling
anomaly appears in quantum mechanics.

	The organization of the paper is as follows. In Section 2 we provide
a general analysis of $N$-particle scale invariant systems in one and higher 
dimensions. Using a suitable separation of variables, the effective    
Hamiltonian in the ``radial'' coordinate is obtained. The 
quantization of the effective Hamiltonian together with its self-adjoint
extension is discussed in Section 3. It is shown that new bound and
scattering states emerge naturally in this analysis. In Section 4, we
discuss the application of this quantization method to several examples. These
include $A_{N-1}$, $B_N$ and $D_N$ Calogero models in one dimension 
and Calogero-Marchioro model in higher dimensions. The restriction 
imposed on the system parameters necessary for the existence of the new 
quantum states is discussed in each case. We conclude the paper in Section 5 
with some discussions.

\sxn{Effective Hamiltonian for Scale Invariant Potentials}

In this Section we show that the Hamiltonian of a class of many-particle
systems with long range and classically scale invariant interactions
can be related to the
Hamiltonian of a single particle with inverse-square interaction on the 
half line. Let us consider a system of $N$ identical particles of mass
$m$ in one dimension whose Hamiltonian in units of $2 m {\hbar}^{- 2} = 1$
is given by
\be
H = \sum_{i=1}^N \left [ - \partial_i^2 + (\partial_i w )^2 - 
\partial_i^2 w \right ],  
\label{p1}
\ee
where $\partial_i$ denotes the derivative with respect to the position of
the $i^{\mathrm th}$ particle. 
The function $w$ appearing in Eqn. (2.1) depends on 
the particle coordinates and is assumed to satisfy the relation
\be
w = - {\mathrm ln} G, 
\label{p3}
\ee
where $G$ is a homogeneous function of degree $d$, i.e.
\be
\sum_{i=1}^N x_i \partial_i G = d  G.
\ee 
Using Eqns. (2.1) - (2.3), we can show that the many-body interaction in $H$
scales in the same way as the kinetic energy term. This implies that there
are no dimensionfull coupling constants in the system, indicating that 
$H$ is classically scale invariant. It may be noted that $O(2,1)$ appears as 
the spectrum generating algebra for $H$ \cite{jackiw1,fubini,susy}.  

We would like to solve the Schr$\ddot{\mathrm o}$dinger's equation
\be
H \psi = E \psi.
\ee
For this purpose, consider an ansatz of the form
\be
\psi = e^{- w} P_k(x) \phi (r),
\ee
where 
$x \equiv (x_1, x_2, \dots, x_N)$ and 
$\phi(r)$ is taken to be a function of the ``radial''
variable $r$ defined by $r^2 = \sum_{i=1}^N x^2_i$. The function  
$P_k (x) $ is a homogeneous polynomial of the particle coordinates
with degree $k \geq 0$, i.e.
\be
\sum_{i=1}^N x_i \partial_i P_k = k P_k,
\ee
satisfying the equation
\be
- \sum_{i=1}^N \partial_i^2 P_k + 
2 \sum_{i=1}^N \partial_i w  \partial_i P_k = 0.
\label{p7}
\ee
Note that for $k=0$, $P_0$ as an
arbitrary constant ( which can be chosen to be unity without any loss of
generality ) is a valid solution of Eqn. (2.7). The solutions of Eqn. (2.7)
for higher values of $k$ are known only for a few one dimensional many-body
systems. If in a given model the solutions of Eqn. (2.7) for nonzero values 
of $k$ are not known, we shall restrict our analysis 
only to the $k = 0$ sector, with $P_0 = 1$.
Substituting Eqn. (2.5) in (2.4) and using Eqns. (2.3), (2.6) and (2.7), we
get
\be
\left [ - \frac{\partial^2 }{\partial r^2} - (1+ 2 \nu )\frac{1}{r}
\frac{\partial }{\partial r} \right ] \phi (r) = E \phi (r),
\ee
where
\be
\nu = \frac{N}{2} - 1 + k + d.
\label{p11}
\ee
Lets us now define the function $\chi(r)$ through the transformation
\be
\phi (r) = r^{-(\frac{1}{2} + \nu)} \chi(r).
\ee
Substituting Eqn. (2.10) in Eqn. (2.8) we get
\be
\tilde{H} \chi (r) \equiv 
\left [ - \frac{\partial^2 }{\partial r^2} + \frac{\nu^2 - \frac{1}{4}}{r^2}
\right ] \chi (r) = E \chi (r).
\ee
The operator $\tilde{H}$ in Eqn. (2.11) defines the effective Hamiltonian 
of the system in the ``radial'' variable.

In deriving Eqn. (2.11) we did not make any assumption about translation
invariance of the system. If the system is translationally invariant, then
both $P_k(x)$ and $w(x)$ obey the same symmetry.
In that case, we can still derive
Eqn. (2.11) after eliminating the center of mass coordinate with $r$ and 
$\nu$ now being given by
\bea
r^2 &=& \frac{1}{2 N} \sum_{i,j=1}^N \left ( x_i - x_j \right
)^2,\\
\nu &=& \frac{N-1}{2} - 1 + k + d.
\label{p8}
\eea
Comparing Eqns. (2.9) and (2.13), we find that these two expressions differ
by a contribution coming from the center of mass of the system.

In this paper we shall discuss both the scattering and bound states sectors 
of the Hamiltonian $H$, or equivalently, of the effective Hamiltonian 
$\tilde{H}$. For the purpose of discussion of the bound states,
we now find the measure for which the eigenfunctions $\chi(r)$ 
would be square integrable. 
Using Eqns. (2.2), (2.5) and (2.10), we get
\be
\psi = G P_k(x) r^{-(\frac{1}{2} + \nu)} \chi(r).
\ee
Note that $r$ takes values in the positive real axis.
For translationally invariant systems, the Hamiltonian $H$ can 
equivalently be described using
$ (N - 1) $ dimensional hyperspherical coordinates with the
radius $r$ and the $(N-2)$ angular variables $\Omega_i$. 
In this case, both $G$ and $P_k (x)$ are translationally invariant
homogeneous function of degree $d$
and $k$ respectively. The wavefunction $\psi(x)$ can thus be factorized as 
$\psi(x) = \xi(r) F(\Omega)$ where
\be
\xi(r) = r^{d + k - \nu -\frac{1}{2}}\chi(r),
\ee
and $F(\Omega)$ is a function of the angular variables $\Omega_i$ alone.
Using the Eqns. (2.13) and (2.15), we see that
\be
\int_0^{\infty} \xi^*(r) \xi(r) r^{(N -2)} dr = \int_0^{\infty}
\chi^*(r) \chi(r) dr,
\ee
which implies that $\chi(r) \in L^2[R^+,dr]$. 
For translationally non-invariant systems, the Hamiltonian $H$ can
be described using $N$ dimensional hyperspherical coordinates. The same line 
of argument as above using Eqn. (2.9) instead of Eqn. (2.13) again gives 
$\chi(r) \in L^2[R^+,dr]$.

We have so far restricted our attention where the space dimension is one.
It can easily be shown that the above derivation is also valid in $D$ space 
dimensions. In this case the particle coordinates would be D
dimensional. Thus substituting $x_i \rightarrow {\bf r}_i$ and 
$\partial_i \rightarrow {\bf \nabla}_i$ in Eqn. (2.1) and following the
subsequent analysis, we get the same effective
Hamiltonian $\tilde{H}$ with $\nu$ being given by 
\be
\nu = \frac{(N-1)D}{2} - 1 + k + d
\label{h1}
\ee
for translationally invariant systems and
\be
\nu = \frac{ND}{2} - 1 + k + d
\label{h2}
\ee
for translationally non-invariant systems. Note that Eqns. (\ref{h1}) and
(\ref{h2}) differ by a contribution coming from the center of mass of the
system.

\sxn{Quantization of the Effective Hamiltonian}

	In this Section we shall discuss the quantization 
of the effective Hamiltonian in Eqn. (2.11) \cite{meetz}. We shall present
the analysis in the scattering sector directly using von Neumann's theory of
deficiency indices \cite{reed} when $\nu^2 > 0$. This will lead to an
alternate derivation of the bound state solutions for this system as well. 
The operator $\tilde{H}$ 
belongs to a general class of objects known as  unbounded linear differential 
operators on a Hilbert space \cite{reed}. We shall first outline some
properties of these operators which are needed for our purpose.

	Let $T$ be an unbounded differential operator acting on a Hilbert
space ${\cal H}$ and let $D(T)$ be the domain of $T$. The inner product 
of two element $\alpha , \beta \in {\cal H}$ is denoted by $(\alpha ,
\beta)$. Let $D(T^*)$ be the set
of $\phi \in {\cal H}$ for which there is a unique $\eta \in {\cal H}$ with
$(T \xi , \phi) = (\xi , \eta )~ \forall~ \xi \in D(T)$. For each such
$\phi \in D(T^*)$, we define $T^* \phi = \eta$. $T^*$ then defines the adjoint
of the operator $T$ and $D(T^*)$ is the corresponding domain of the adjoint.
The operator $T$ is called symmetric or Hermitian iff $(T \phi, \eta) = 
(\phi, T \eta) ~ \forall ~ \phi, \eta \in D(T)$. The operator $T$ is called 
self-adjoint iff $T = T^*$ {\it and} $D(T) = D(T^*)$. 

We now state the criterion to determine if a symmetric operator $T$ is
self-adjoint. For this purpose let us define the deficiency subspaces 
$K_{\pm} \equiv {\rm Ker}(i \mp T^*)$ and the 
deficiency indices $n_{\pm}(T) \equiv
{\rm dim} [K_{\pm}]$. Then $T$ falls in one of the following categories:\\
1) $T$ is (essentially) self-adjoint iff
$( n_+ , n_- ) = (0,0)$.\\
2) $T$ has self-adjoint extensions iff $n_+ = n_-$. There is a one-to-one
correspondence between self-adjoint extensions of $T$ and unitary maps
from $K_+$ into $K_-$. \\
3) If $n_+ \neq n_-$, then $T$ has no
self-adjoint extensions.

We now return to the discussion of the effective Hamiltonian $\tilde{H}$.
This is an unbounded differential operator defined in    
$R^+ $. $\tilde{H}$ is a symmetric operator on the domain
$D(\tilde{H}) \equiv \{\phi (0) = \phi^{\prime} (0) = 0,~
\phi,~ \phi^{\prime}~  {\rm absolutely~ continuous},~ \phi \in {\rm L}^2(dx)
\} $.
We would next like to determine if $\tilde{H}$ is self-adjoint, for which 
we treat the cases $\nu \neq 0$ and $\nu = 0$ separately. 

\noindent
{\bf 3.1}~ {\boldmath ${\nu \neq 0}$}

The deficiency indices $n_{\pm}$ are determined by the number of
square-integrable solutions of the equations
\be
\tilde{H}^* \phi_{\pm} = \pm i \phi_{\pm},
\ee
respectively, where $\tilde{H}^*$ is the adjoint of $\tilde{H}$. Note that 
$\tilde{H}^*$ is given by the same differential operator as $\tilde{H}$.
From dimensional considerations we see that the r.h.s. of Eqn. (3.1) 
should be multiplied with a constant with dimension of ${\rm length}^{-2}$.
We shall henceforth choose the magnitude of this constant to be unity 
by appropriate 
choice of units. The solutions of Eqn. (3.1) are given by
\bea
\phi_+ (r) &=& r^{\frac{1}{2}}H^{(1)}_\nu (re^{i \frac{ \pi}{4}}),\\
\phi_- (r) &=& r^{\frac{1}{2}}H^{(2)}_\nu (re^{-i \frac{ \pi}{4}}),
\eea
where $H_\nu$'s are Hankel functions \cite{abr}. 
The functions $\phi_{\pm}$ are bounded as $r \rightarrow \infty $. When 
$r \rightarrow 0$, they behave as 
\bea
\phi_+ (r) &\rightarrow & \frac{i}{{\rm sin}\nu \pi}
\left [ \frac{r^{\nu + \frac{1}{2}}}{2^\nu} 
 \frac{ e^{ - i  
\frac{3 \nu \pi}{4}}}{\Gamma (1 + \nu)}
- \frac{r^{- \nu + \frac{1}{2}}}{2^{- \nu}}
\frac{e^{-i \frac{ \nu \pi}{4}} }{\Gamma (1 -
\nu)} 
\right ],\\
\phi_- (r) &\rightarrow & \frac{i}{{\rm sin}\nu \pi}
\left [- \frac{r^{\nu + \frac{1}{2}}}{2^\nu}
 \frac{ e^{  i  
\frac{3 \nu \pi}{4}}}{\Gamma (1 + \nu)}
+ \frac{r^{- \nu + \frac{1}{2}}}{2^{- \nu}}  
\frac{e^{i \frac{ \nu \pi}{4}} }{\Gamma (1 -
\nu)}   
\right ].
\eea
We see that $\phi_{\pm}$ are not square integrable when 
${\nu}^2 \geq 1$. In this case 
$\tilde{H}$ has deficiency indices (0,0) and is
(essentially) self-adjoint on the domain $D(\tilde{H})$ \cite{hut}.
On the other hand, both $\phi_{\pm}$ are square
integrable when either $ -1 < \nu < 0$ or $ 0 < \nu < 1$. We therefore see 
that for any value of $\nu$ in these ranges, 
$\tilde{H}$ has deficiency indices $ (1,1)$. In this case, 
$\tilde{H}$ is not self-adjoint on the 
domain $D(\tilde{H})$ but admits self-adjoint extensions.
The deficiency
subspaces  $K_{\pm}$ in this case are one dimensional and are spanned by
the functions $\phi_{\pm}$.
The unitary maps from $K_+$ into $K_-$ are parameterized by $e^{i z}$ where
$z \in R$ (mod $2 \pi$).
The operator $\tilde{H}$ is self-adjoint in the domain
$D_z(\tilde{H}) = D(\tilde{H}) \oplus \{ a(\phi_+ (r) + e^{i z} \phi_- (r))
\}$ where $a$ is an arbitrary complex number \cite{reed,hut}. 

We would now like to obtain the spectrum of $\tilde{H}$ in the
domain $D_z(\tilde{H})$.
We start with the discussion of the scattering states given 
by the 
positive energy solutions of Hamiltonian $\tilde{H}$. For that we set $E=q^2$, 
where $q$ is a real positive parameter. The general solution
of Eqn. (2.11) can be written as 
\be
\chi (r) = r^{1\over 2} \left [ a(q) J_\nu (qr) - b(q) J_{- \nu} (qr) \right
]
\ee
where $a(q)$ and $b(q)$ are two as yet undetermined coefficients. $J_{\nu}$
in Eqn. (3.6) refers to the Bessel function of order $\nu$ \cite{abr}. 
Note that in the limit $r \rightarrow 0$,
\be
\phi_{+}(r) + e^{i z} \phi_{-}(r) \rightarrow \frac{i}{{\rm sin}\nu \pi}
\left [ \frac{r^{\nu + \frac{1}{2}}}{2^\nu} \frac{( e^{ - i  
\frac{3 \nu \pi}{4}} - e^{i (z +  \frac{3 \nu \pi}{4})})}{\Gamma (1 + \nu)}
+ \frac{r^{- \nu + \frac{1}{2}}}{2^{- \nu}}
\frac{( e^{ i
( z +  \frac{\nu \pi}{4} ) } - e^{-i \frac{ \nu \pi}{4}}) }{\Gamma (1 -
\nu)}
\right ]
\ee     
and     
\be     
\chi (r) \rightarrow   \frac{r^{\nu + \frac{1}{2}}}{2^\nu}
\frac{a(q)q^\nu}{\Gamma (1 + \nu)} 
- \frac{r^{- \nu + \frac{1}{2}}}{2^{- \nu}}
\frac{b(q)q^{-\nu}}{\Gamma (1 -\nu)}  \, . 
\ee
If $\chi (r) \in D_z(\tilde{H})$, then the coefficients of
of $r^{\nu + \frac{1}{2}}$ and $r^{- \nu + \frac{1}{2}}$
 in Eqns. (3.7) and (3.8) must match. Comparing these coefficients we get
\be
\frac{a(q)}{b(q)} =  \frac{{\rm sin}(\frac{z}{2} + 3 \pi \frac{\nu}{4})}
{{\rm sin}(\frac{z}{2} + \pi \frac{\nu}{4})} q^{-2\nu }  \, .
\ee
Next we calculate the $S$-matrix and the associated phase shift.
In the limit $r \rightarrow \infty$,
the leading term in the asymptotic expansion of $\chi (r)$ 
in Eqn. (3.6) is given
by 
\be
\chi(r) \rightarrow
\frac{1}{\sqrt{2\pi q}}e^{iqr}
\left [ a(q) e^{-i (\nu + {1\over 2}) {\pi \over 2}}  -
b(q) e^{i (\nu - {1\over 2}) {\pi \over 2}}  \right ]  +
\frac{1}{\sqrt{2\pi q}}e^{-iqr} 
\left [ a(q) e^{i (\nu + {1\over 2}) {\pi \over 2}}  -
b(q) e^{-i (\nu - {1\over 2}) {\pi \over 2}}  \right ].
\ee
By dividing the coefficient of outgoing wave ($e^{iqr}$) by that
of the incoming wave ($e^{-iqr}$), we obtain the 
the $S$-matrix and phase shift $\delta (q)$ as
\be 
S(q) \equiv e^{2 i \delta(q)} =
\frac{ a(q) e^{-i (\nu + {1\over 2}) {\pi \over 2}}  -
b(q) e^{i (\nu - {1\over 2}) {\pi \over 2}} }
{ a(q) e^{i (\nu + {1\over 2}) {\pi \over 2}}  -
b(q) e^{-i (\nu - {1\over 2}) {\pi \over 2}} }  \, .
\ee
Finally, using Eqns. (3.9) and (3.11), we get
\be 
S(q) = e^{2 i \delta(q)} =
\frac{  
q^{-\nu} \sin ( \frac{z}{2} + 3 \pi \frac{\nu}{4} )
  e^{-i (\nu + {1\over 2}) {\pi \over 2}}  - 
q^\nu \sin ( \frac{z}{2} + \pi \frac{\nu}{4} )
e^{i (\nu - {1\over 2}) {\pi \over 2}} }
{  q^{-\nu} \sin ( \frac{z}{2} + 3 \pi \frac{\nu}{4} )
  e^{i (\nu + {1\over 2}) {\pi \over 2}}  - 
q^\nu \sin ( \frac{z}{2} + \pi \frac{\nu}{4} )
e^{ - i (\nu - {1\over 2}) {\pi \over 2}} }   \, .
\ee

Let us now consider the bound state solutions of Eqn. (2.11).
For any given value of $\nu $ in the allowed range
the $S$-matrix
in Eqn. (3.12) has a pole on the positive imaginary axis of the complex    
$q$-plane. Such a pole indicates the existence 
of a bound state for the effective Hamiltonian $\tilde{H}$. By taking
 $q= i\rho $ as the pole for the $S$-matrix in Eqn. (3.12),  one can 
easily derive the corresponding bound state energy $E = - \rho^2$ as 
\be
E  
 = - \left [ \frac{{\rm sin}(\frac{z}{2} + 3 \pi \frac{\nu}{4})}
{{\rm sin}(\frac{z}{2} + \pi \frac{\nu}{4})} \right ]^{\frac{1}{\nu}}.
\ee
Thus we see that for a given value of $\nu$ within the allowed range,
$\tilde{H}$ admits a single bound state with energy given by Eqn. (3.13).
It may be noted that for a fixed $\nu$, the bound state exists only for
those values of $z$ such that the quantity 
$\frac{{\rm sin}(\frac{z}{2} + 3 \pi \frac{\nu}{4})}
{{\rm sin}(\frac{z}{2} + \pi \frac{\nu}{4})}$ in Eqn. (3.13) 
is positive \cite{meetz}.
Using Eqns. (3.6), (3.9) and (3.13), we see that the 
the corresponding bound state eigenfunction is given by
\be
\chi (r) = B r^{\frac{1}{2}} H^{(1)}_\nu(i \sqrt{|E|} r),
\ee
where $B$ is the normalization constant. 
It may be noted that the expression of the bound state
wavefunction for $0 < \nu < 1$ has been discussed in the 
literature \cite{meetz,trg,kumar1} and agrees with Eqn. (3.14), when 
restricted to the same range of $\nu$.  

We would like to remind the reader that when $\nu^2 \geq 1$,
$\tilde{H}$ is essentially self-adjoint in the domain
$D(\tilde{H})$. In this case, the scattering state solutions are given by
\be
\chi (r) = a(q) r^{1\over 2} J_\nu (qr),
\ee
which is same as Eqn. (3.6) without the term proportional to $J_{-\nu}
(qr)$. No bound states exist in this case. 

\noindent
{\bf 3.2}~ {\boldmath ${\nu = 0}$}

In this case, the solutions of the equations
\be
\tilde{H}^* \psi_{\pm} = \pm i \psi_{\pm}
\ee
are given by
\bea
\psi_+ (r) &=& r^{\frac{1}{2}}H^{(1)}_0 (re^{i \frac{ \pi}{4}}),\\
\psi_- (r) &=& r^{\frac{1}{2}}H^{(2)}_0 (re^{-i \frac{ \pi}{4}})
\eea
respectively. $\psi_{\pm}$ are bounded functions as $r \rightarrow \infty$. 
In order to find their behaviour for small $r$, 
we first note that in the limit $r \rightarrow 0$,
\bea
J_0(r) &=& 1 + {\cal O} \left ( r^2 \right )\\
N_0(r) &=& \frac{2}{\pi} \left [ \gamma - {\rm {ln}}2 + {\rm {ln}}r \right ]
           + {\cal O} \left ( r^2 {\mathrm ln} r \right )
\eea
where $\gamma$ is Euler's constant and $N_0$ is the Neumann function
\cite{abr}.
Using Eqns. (3.17), (3.19) and (3.20) we see that when $r \rightarrow 0$,
$\psi_+ (r)$ behaves as 
\be 
\psi_+ (r) \rightarrow \frac{2 i}{\pi} r^{\frac{1}{2}} {\rm {ln}} r
             + r^{\frac{1}{2}} \left [ \frac{1}{2} + \frac{2 i}{\pi}
                ( \gamma - {\rm {ln}}2 ) \right ].
\ee
In the same limit, $\psi_- (r)$ behaves as
\be
\psi_- (r) \rightarrow -\frac{2 i}{\pi} r^{\frac{1}{2}} {\rm {ln}} r
             + r^{\frac{1}{2}} \left [ \frac{1}{2} - \frac{2 i}{\pi}
                ( \gamma - {\rm {ln}}2 ) \right ].
\ee
From Eqns. (3.21) and (3.22) we see that 
both $\psi_{\pm} (r)$ are square integrable functions \cite{moretti}.
$\tilde{H}$ therefore has deficiency indices (1,1) and the corresponding 
deficiency subspaces $ K_{\pm} $ are again 1-dimensional, spanned by
the functions $\psi_{\pm} (r)$.
As before, the operator ${\tilde H}$ is not self-adjoint on
$D({\tilde H})$ but admits a one-parameter family of self-adjoint extensions
labeled by $e^{i z}$ where $z \in R$.  
It is self-adjoint in the domain
$D_z(\tilde{H})$ which contains all the elements of $D(\tilde{H})$ together  
with elements of the form $a(\psi_+ (r) + e^{i z} \psi_- (r)) $ where $a$ is
an arbitrary complex number.
\cite{reed,hut}.

The scattering states associated with positive energy solutions
 of Eqn.(2.11) are given by 
\be
\chi (r) = r^{1\over 2} \left [ a(q) J_0(qr) - b(q) N_0(qr) \right ],
\ee
where $E=q^2$ and $a(q), ~b(q)$ are two as yet undetermined coefficients. 
In order to find the ratio of these two coefficients, 
as before we use the fact that if ${\tilde H}$ has to be
self-adjoint, the eigenfunction $\chi (r)$ must belong to the domain
$D_z({\tilde H})$.
In the limit when $r \rightarrow 0$, we have
\be 
\chi(r) \rightarrow - \frac{2b(q)}{\pi} r^{1\over 2} \ln r + r^{1\over 2} 
\left[a(q)-{2b(q)\over \pi} \ln q + 
{2b(q)\over \pi} (\ln 2 - \gamma ) \right].
\ee
We now compare the coefficients of $r^{1\over 2}$ and 
$r^{1\over 2}{\rm {ln}} r$ in Eqn. (3.24) 
with those appearing in the expression of
$e^{i \frac{z}{2}} \psi_- (r) + e^{-i\frac{z}{2}}\psi_+ (r)$. Comparing the
coefficients of $r^{1\over 2}{\rm {ln}} r$ we find
\be
b(q) = - 2 {\rm sin} \frac{z}{2}.
\ee
Comparing the coefficients of $r^{1\over 2}$ and using Eqn. (3.25) we obtain
\be
 a(q) - {2b(q)\over \pi} \ln q  
 =  {\rm cos} \frac{z}{2}.
\ee
Using Eqns. (3.25) and (3.26) and assuming that $z \neq 0$, 
we finally obtain
\be
{a(q)\over b(q)} = \frac{2}{\pi} \ln q  - {1\over 2} {\rm {cot}}
\frac{z}{2}.
\ee
Thus we find that for any generic value of the self-adjoint parameter 
$z \neq 0$,
the ratio of $a(q)$ and $b(q)$ depends on the momentum 
$q$. We shall comment on the $z = 0$ case at the end of this Section.

For calculating the $S$-matrix and phase shift for the
above mentioned scattering process, we consider
the leading term in the asymptotic expansion of $\chi (r)$ 
of Eqn. (3.23) 
 at $r \rightarrow \infty$ limit. This gives 
\be
\chi(r) \rightarrow
\frac{1}{\sqrt{2\pi q}}e^{iqr} e^{-i{\pi \over 4}}
\left [ a(q) + i b(q)  \right ]  +
\frac{1}{\sqrt{2\pi q}}e^{-iqr} e^{i{\pi \over 4}}
 \left [ a(q)  - i b(q)  \right ].
\ee
By dividing the coefficient of outgoing wave ($e^{iqr}$) by that
of the incoming wave ($e^{-iqr}$) and using Eqn. (3.28), we obtain the 
the $S$-matrix and phase shift as
\be
S(q) \equiv e^{2 i \delta(q)} = e^{-i { \pi \over 2}}
{ \frac{2}{\pi} \ln q - \frac{1}{2} \cot \frac{z}{2}  + i \over 
 \frac{2}{\pi} \ln q - \frac{1}{2} \cot \frac{z}{2}  - i }.
\ee
When $z \neq 0$, the $S$-matrix in Eqn. (3.29) again
 has a single pole on the positive imaginary axis 
of the complex $q$-plane. 
Such  a pole naturally  indicates the existence 
of a bound states for the effective Hamiltonian
${\tilde{H}}$. 
By taking $q = i\mu $ as the pole for the $S$-matrix 
in Eqn. (3.29),  we obtain 
 the corresponding bound state energy $E = - \mu^2$ as 
\be
E=  
- {\rm {exp}} \left [ \frac{\pi}{2}  {\rm {cot}} 
\frac{z}{2} \right ].
\ee
 The corresponding eigenfunction is given by 
\be
\chi (r) =  C r^{\frac{1}{2}} K_0 \left ( \sqrt{|E|} r \right ) 
= C \frac{i \pi}{2} r^{\frac{1}{2}} 
H_0^{(1)} \left (i \sqrt{|E|} r \right ),
\ee
where $C$ is a constant and $K_0$ is the modified Bessel function
\cite{abr}.

	This concludes our discussion regarding the quantization of the
effective Hamiltonian $\tilde{H}$. We end this Section with the following 
general observations.

\noindent
1)	We have seen that for a 
given value of $\nu$, each value of the parameter $z$  (mod $2 \pi$) produces 
a different spectrum of $\tilde{H}$. The spectrum of $\tilde{H}$, and
correspondingly the parameter space of the quantum theory 
is thus characterized by the pair $(\nu,z)$. Let us assume for
the moment that $\nu \neq 0$. Consider now two pairs of parameters $(\nu,z)$
and $(\nu^{\prime},z^{\prime})$ corresponding to two different quantum
theories. It is easily seen that the spectrum of these two theories 
are identical if
$\nu^{\prime} = -\nu$ and $z^{\prime} = z + 2 \pi \nu$.
We therefore have an
equivalence relation $(\nu,z) \sim (-\nu, z + 2 \pi \nu)$ on the parameter
space of $\tilde{H}$. The 
transformation $(\nu,z) \rightarrow (\nu^{\prime},z^{\prime})$ relating
two different quantum theories is analogous to the 
duality symmetry in this system. 
In the case when $\nu = 0$, we automatically have
$(\nu,z) = (\nu^{\prime},z^{\prime})$ for all values of $z$. We can thus say
that the pair $(\nu = 0,z)$ defines the self-dual points in the parameter
space.

\noindent
2) As mentioned before, the classical system that we started with possesses 
scaling symmetry. However, in the presence of the 
self-adjoint extensions, the system admits bound state(s) and 
the phase shifts in the scattering sector depend explicitly on the 
momentum. These are indicative of the breakdown of the scaling symmetry at the
quantum level. We first analyze this issue when $-1 < \nu \neq 0 < 1$. 
Let us consider the action of the
scaling operator $\Lambda = \frac{-i}{2} (r \frac{d}{dr} + \frac{d}{dr} r)$
on an element $\phi(r) = \phi_{+}(r) + e^{i z} \phi_{-}(r)
\in  D_z({\tilde H})$. In the limit $r \rightarrow 0$, we have
\be
\Lambda \phi(r) \rightarrow
\frac{1}{{\rm sin}\nu \pi}
\left [ (1 + \nu) \frac{r^{\nu + \frac{1}{2}}}{2^\nu} \frac{ ( e^{ - i    
\frac{3 \nu \pi}{4}} - e^{i (z +  \frac{3 \nu \pi}{4})})}{\Gamma (1 + \nu)}
+ (1 - \nu) \frac{r^{- \nu + \frac{1}{2}}}{2^{- \nu}}
\frac{ ( e^{ i
( z +  \frac{\nu \pi}{4} ) } - e^{-i \frac{ \nu \pi}{4}}) }{\Gamma (1 -
\nu)}    
\right ].
\ee   
In order for $\Lambda \phi(r) \in D_z({\tilde H})$, we must have
$\Lambda \phi(r) \sim C \phi(r)$ where $C$ is a constant. However,
the two terms on the r.h.s. of Eqn. (3.32) are multiplied by two different
factors , i.e. $(1 + \nu )$ and $(1 - \nu )$. Due to the presence of these
different multiplying factors, we see that
$\Lambda \phi(r)$ in general does not belong
to $D_z({\tilde H})$. Scale invariance is thus broken at the quantum level
for
generic values of $z$. However, from Eqn. (3.32) it is clear that
for special choice of $z = -\frac{ \nu \pi}{2}$, $\Lambda \phi(r) \in
D_z({\tilde H})$ and the scaling symmetry is recovered \cite{trg}. In
addition, we find that the scaling symmetry is also 
preserved at the quantum level when  $ z = -\frac{ 3 \nu \pi}{2}$.
For these choices of $z$, 
the bound states do not exist and the $S$ matrix becomes independent of the
momentum. The scaling symmetry thus is anomalously broken due to the
quantization for generic values of $z$. In the case when $\nu =0$, a similar
analysis as above again shows that the self-adjoint extension generically 
breaks the scaling symmetry. For this case, the scale invariance can be 
recovered at the quantum level only for $z = 0$.

\sxn{Examples}

We have shown in the previous Section that the effective Hamiltonian 
$\tilde{H}$, and thus the Hamiltonian $H$, admits self-adjoint extensions 
when $ - 1 < \nu < 1$. As discussed before, $\nu$ is a function of
several system parameters. The condition on $\nu$ implies that $\tilde{H}$
admits self-adjoint extensions only if the system parameters are suitably
constrained. In this Section we shall consider several models with suitable
ranges of parameters, such that the corresponding Hamiltonians admit 
self-adjoint extensions. The new bound and scattering states will be
obtained by substituting the solutions of eigenvalue Eqn. (2.11) into Eqn.
(2.14). As mentioned before, the effective Hamiltonian is essentially
self-adjoint when $\nu^2 \geq 1$. This is the case considered by Calogero in
Ref. \cite{calo1} for the $A_{N-1}$ model. In the examples discussed below
we shall not consider the range $\nu^2 \geq 1$. 

\noindent
{\bf 4.1 {\boldmath $A_{N-1}$} Calogero Model}

The Hamiltonian of the $A_{N-1}$ Calogero Model without the harmonic term
is given by 
\be
H_{A_{N-1}} = - \sum_i \frac{d^2}{d x_i^2} + (a^2 - \frac{1}{4} )
\sum_{i \neq j} \left ( x_i - x_j \right )^{-2},
\label{e1}
\ee
where $i,j = 1, 2, \cdots, N$.
This Hamiltonian can be obtained from Eqns. (2.1) and (2.2) by choosing 
$G$ as
\be
G = \prod_{i < j} ( x_i - x_j )^{a + \frac{1}{2}},
\ee
which is a homogeneous function of degree
$d= (a+\frac{1}{2}) N (N-1)/2$. The Hamiltonian $H_{A_{N-1}}$ is
translationally invariant and the parameter $\nu$ is determined from
Eqn. (2.13) as 
\be
\nu = k + \left ( a + \frac{1}{2} \right ) \frac{N ( N- 1 )}{2} + 
\frac{ N - 3}{2}.
\ee
\noindent For this model, it 
is known that Eq. (\ref{p7}) determining $P_k(x)$ is exactly solvable for
any $k \geq 0 $ \cite{calo1,pr}. Below we shall discuss both bound and
scattering state sectors of the Hamiltonian $H_{A_{N-1}}$ for 
arbitrary $ k \geq 0$.

The bound state sector of this system was analyzed in Ref. \cite{kumar1}
under the assumption that $a + \frac{1}{2} \geq 0$. 
In this case, the wavefunction vanishes in the limit when
any two particle coordinates coincide, though
the corresponding current might show a divergent behaviour.
Such a boundary condition on the wavefunction is quite similar to what one 
encounters in the case of $\delta$-function Bose gas with infinitely large 
value of two-body coupling constant \cite{poly,ha}.
This type of  boundary condition allows one to construct 
continuous eigenfunctions on the whole of 
 configuration space by first solving the eigenvalue problem
for a definite ordering of particles e.g. $x_1 \geq x_2 \geq \cdots \geq
x_N$ and then smoothly extending it to the rest of the configuration 
space using permutation symmetry associated with
identical particles.  It is obvious from Eqn.
(4.3) that for $ a + \frac{1}{2} \geq 0 $ and $N \geq 3$, $\nu$ is a positive
definite quantity. Thus, the allowed ranges
of $\nu$ for $H_{A_{N-1}}$ to have bound states is further restricted to
$ 0 < \nu < 1 $. Analyzing Eqn. (4.3) with the constraints on
$a+\frac{1}{2}$ and $\nu$ discussed above, it was shown in Ref.
\cite{kumar1} 
that the $A_{N-1}$ Calogero model admits negative energy bound states only
for
$N=3,4$ and $k=0$.

The bound state wavefunctions of $H_{A_{N-1}}$ obtained in Ref.
\cite{kumar1} are normalizable, although as $r \rightarrow 0$,
they have a singularity of the form $r^{- {\nu } - \frac{N-3}{2} }$. 
Note that the singularity at $r=0$ corresponds to the case
where all the particles
coincide at the same point. This is different from the singularity  
arising from the coincidence of any two particles, which is avoided 
by imposing the
constraint $a + \frac{1}{2} \geq 0$. However, from the viewpoint of
self-adjoint extensions alone, there are no reasons a priori to 
distinguish between the singularity at $r\rightarrow 0 $ limit and the
singularity arising from the coincidence of any two particle coordinates.
In view of this, we now let both these types of  singularities to appear 
in the wavefunction. It may be noted that for both bound and scattering 
state solutions, the angular part of the total
wavefunction must be square integrable. The angular part of the wavefunction
receives a contribution from the factor $G$ in Eqn. (4.2), which is singular
when $ a + \frac{1}{2} < 0$. Requiring square 
integrability of the angular part of the wavefunction puts the restriction  
that $ a + \frac{1}{2} > - \frac{1}{2}$. 
A negative value for the parameter $a + \frac{1}{2} $ in this range
leads to a singularity in the wavefunction 
resulting from the coincidence of any
two particle coordinates. This restricts the range of a continuous
eigenfunction within a region of configuration space corresponding to
definite ordering of particles like $x_1 \geq x_2 \geq \cdots \geq x_N$.
Using permutation symmetry, such an eigenfunction can be extended to the 
whole of phase space, although not in a smooth fashion.
The bound state wavefunctions of $H_{A_{N-1}}$ thus obtained are completely
normalizable for $ a + \frac{1}{2} > - \frac{1}{2}$.

We now analyze the case where $N$ is fixed while $ a +
\frac{1}{2}$ is allowed to vary subject to the restriction
that $ a + \frac{1}{2} > - \frac{1}{2}$. We note that since $-1 < \nu < 1$,
we must have,
\be
- \frac{ N - 1 + 2 k}{N ( N-1)} < a + \frac{1}{2} < -
\frac{ N - 1 + 2 k}{N ( N-1)} + \frac{4}{N (N-1)}.
\label{pp}
\ee
\noindent 
Imposing the condition that the upper
bound on $a + \frac{1}{2}$ in Eqn. (\ref{pp}) should be greater than
$-\frac{1}{2}$,
we find that $k$ is restricted as 
\be
k < \frac{1}{4} \left ( N^2 - 3 N + 10 \right ).
\ee
\noindent Thus, the allowed values of $k$ are $0, 1, 2, \dots,
K \equiv \{(N^2 - 3 N + 10)/4 \}$. 
The symbol $\{x \}$ denotes the integral part of $x$ if $x$ is non-integer
and is equal to $x-1$ for integer $x$. 
For a fixed value of $N$, the self-adjoint extension is
admissible when the coupling constant lies within the range
\be
 -\frac{1}{2} < a+\frac{1}{2} < \frac{5 -N}{N(N-1)}.
\ee
Note that the upper bound in Eqn. (4.6) is negative for $N > 5$. This
implies that the interaction in Eqn. (4.1) is repulsive for $N > 5$.

We have seen in the previous paragraph that for fixed value of $N$ and
variable coupling, the self-adjoint extension and,
hence, new quantum states are admissible for several values of $k$.
We now ask as to what are the allowed values of $k$ when both
$N$ and $a+\frac{1}{2}$ are kept fixed. To this end, we introduce a quantity
$\beta (a,N) = ( a + \frac{1}{2} ) N ( N- 1 )/2 + (N - 3)/2$ so that 
$\nu = k + \beta (a,N)$. Note that for fixed values of $a$ and $N$, $\beta$
is fixed. Moreover, since $a+\frac{1}{2} > - \frac{1}{2}$ ,
$\beta$ is bounded from below, $\beta > - \beta_0 \equiv 
- \frac{1}{4} (N^2 - 3 N + 6)$. It is now easy to see that with
$-1 < \nu < 1$ and $k \geq 0$, 
number of allowed values of $k$ for fixed  $\beta$ is at most 2. 
In order to find the specific values of $k$, we consider the cases $\nu \neq
0$ and $\nu = 0$ separately.

\noindent {\bf 4.1.1}~ {\boldmath ${\nu \neq 0}$}

We now use the constraint, $- 1 < \nu\neq 0 < 1$, and present our results in
the table below.
\def\multic{ \multicolumn{1}{c|} }
\def\bultic{ \multicolumn{2}{c|} }
\def\multboth{ \multicolumn{1}{|c|} }

            \begin{center}
            \begin{tabular}{|p{1.8in}|p{2.2in}|p{0.9in}|}
            \hline
            \multboth{ {\sl {\rm {Bounds on}} $\beta$}} &
          \multic{ {\sl {\rm {Ranges of}} $a+\frac{1}{2}$} }&
           \multic{ {\sl {\rm {Allowed values of}} $k$ }}\\
            \hline
&&\\
\hspace{0.6in} $ \beta \geq 1$ & \hspace{0.65in} $ a + \frac{1}{2} \geq
\frac{5 - N}{N(N-1)}$ & $\hspace{0.6in} \times$\\
&&\\
\hspace{0.45in} $0 < \beta < 1$ & \hspace{0.25in} $ \frac{3 - N}{N(N-1)}
< a + \frac{1}{2} <
\frac{5 - N}{N(N-1)}$ & \hspace{0.6in} 0\\
&&\\
\hspace{0.45in} $- 1 < \beta < 0 $& \hspace{0.35in} $ - \frac{1}{N} < a +
\frac{1}{2} <
\frac{3 - N}{N(N-1)}$ & \hspace{0.5in} 0, 1\\
&&\\
$ \hspace{0.45in} -2 < \beta < - 1$ & $ \hspace{0.35in} -\frac{N+1}{N(N-1)}
< a + \frac{1}{2} < - \frac{1}{N}$& \hspace{0.5in} 1, 2\\
&&\\
\hspace{1in} $\cdot$ & \hspace{1.2in} $\cdot$ &
\hspace{0.6in} $ \cdot$\\
&&\\
$ \hspace{0.3in} - (l+1) < \beta < -l $& $ \hspace{0.2in} -
\frac{N+2 l -1}{N(N-1)} < a + \frac{1}{2} <
- \frac{N+ 2 l -3}{N(N-1)}$ & \hspace{0.4in} $l, l+1$\\
&&\\
\hspace{1in} $\cdot$ & \hspace{1.2in} $\cdot$ &
\hspace{.6in}$ \cdot$\\
&&\\
$-(K-1)<\beta <-(K-2)$&
$ - \frac{N+2 K -5}{N(N-1)} < a + \frac{1}{2} < -
\frac{N + 2 K -7}{N(N-1)}$ & $ \hspace{0.1in}
K-2, K-1$\\
&&\\
$\hspace{0.15in} - \beta_0 < \beta < - (K-1) $&
$ \hspace{0.3in} - \frac{1}{2} < a + \frac{1}{2} < -
\frac{N + 2 K -5}{N(N-1)}$ &$ \hspace{0.3in}
K-1, K$\\
&&\\
\hline
\end{tabular}
\end{center}
\noindent 
It is curious to observe how only two successive values of $k$ are
allowed within a specific range of $a+\frac{1}{2}$, although any lower value
than this $k$ within the same range is not allowed.

When $\nu \neq 0$, using Eqns. (2.14), (3.14) and (4.2), we see that
the bound state wavefunction for $H_{A_{N-1}}$ is obtained as 
\be
\psi = B \prod_{i < j} ( x_i - x_j )^{a + \frac{1}{2}} 
P_k(x) r^{-\nu} H_\nu^{(1)}(i \sqrt{|E|}r).
\ee
The corresponding bound state energy $E$ is given in Eqn. (3.13).
Note that for fixed values of $N$, $a$ and $z$, the system may admit two 
bound states corresponding to two possible values of $k$.  
Similarly, the scattering state solutions of $H_{A_{N-1}}$ 
obtained by using Eqns. (2.14), (3.6) and (4.2) are given by 
\be
\psi = \prod_{i < j} ( x_i - x_j )^{a + \frac{1}{2}}
P_k (x) r^{-\nu} \left [ 
a(q)  J_\nu (qr) - b(q)  J_{- \nu} (qr) \right ],
\ee
where $\frac{a(q)}{b(q)}$ is given in Eqn. (3.9).

\noindent {\bf 4.1.2}~ {\boldmath ${\nu = 0}$}\\

We now discuss the case for which $\nu=0$, or equivalently $\beta = - k$.
It follows from Eqn.(4.3) that $ a+\frac{1}{2} =
\frac{3-N-2k}{N(N-1)}$. Since $a +\frac{1}{2} > - \frac{1}{2}$, the
allowed values of $k$ are $0,1, \dots, \{(N^2 - 3 N + 6)/4 \}$. 
In this case, using Eqns. (2.14), (3.31) and (4.2) we see that
the bound state wavefunction for the Hamiltonian
$H_{A_{N-1}}$ is given by
\be
\psi = C \frac{i \pi}{2} \prod_{i < j} ( x_i - x_j )^{a + \frac{1}{2}}
P_k(x) r^{-\nu}
H_0^{(1)} \left (i \sqrt{|E|} r \right ),
\ee
where the bound state energy $E$ are given in  Eqn. (3.30).
Similarly, the scattering state solutions in this case are 
obtained using Eqns. (2.14, (3.23) and (4.2) as 
\be
\psi = \prod_{i < j} ( x_i - x_j )^{a + \frac{1}{2}} P_k(x) r^{-\nu}
\left [ a(q) J_0(qr) - b(q) N_0(qr) \right ],
\ee
where ${a(q)\over b(q)}$ is given in Eqn. (3.27).

We end this Section with the following observations:\\
(i) Self-adjoint extension and, hence, the existence of the new quantum states
is admissible in the $A_{N-1}$ Calogero model for arbitrary $N$.\\
(ii) The self-adjoint extension for $ k > 0$ is remarkable
in the following sense. If a Hamiltonian admits self-adjoint extension,
it usually occurs in the zero 
angular momentum sector of the Hamiltonian. We are not aware of any
counter-example to this fact. For the $H_{A_{N-1}}$, $k$ can be related
to the angular-momentum eigenvalue of the Hamiltonian\cite{gamb}. Considering
the case $k \neq 0$ amounts to studying the model in the 
non-zero angular momentum sector of the model. Thus, we provide the first
example in the literature where self-adjoint extension 
is admissible in non-zero angular momentum
sector of a Hamiltonian.

\noindent
{\bf 4.2 {\boldmath $B_N$} Calogero model}

The Hamiltonian of the $B_N$ Calogero model without the harmonic term
is given by
\bea
H_{B_N} & = & - \sum_{i=1}^N \frac{d^2}{d x_i^2} + g_1 (g_1 -1 )
\sum_{i \neq j} \left [ (x_i - x_j)^{-2} + (x_i + x_j)^{-2} \right ]
\nonumber\\
&& + g_2 (g_2 -1 ) \sum_{i=1}^N x_i^{-2}.
\label{e6}
\eea
\noindent 
The Hamiltonian in Eqn. (4.11) can  
be obtained from Eqns. (\ref{p1}) and (\ref{p3})
by choosing $G$ as
\be
G = \prod_{i < j} (x_i^2 - x_j^2 )^{g_1} \prod_s x_s^{g_2},
\label{e7}
\ee
\noindent where $i,j,s=1,2, \cdots, N$. $G$ in Eqn. (4.12) 
is a homogeneous function of degree $d=  g_1 N (N-1) + g_2
N$. It is known that for the $B_{N}$ Calogero model, Eqn. (2.7) 
admits solutions only for even $k \geq0$ \cite{pr}. 
Moreover, the Hamiltonian $H_{B_N}$
is not translationally invariant. Consequently, $\nu$ is determined
from Eqn. (\ref{p11}) as 
\be
\nu = \frac{N}{2} -1 + g_1 N (N-1) + g_2 N + 2 k,
\label{e8}
\ee
where $k=0, 1, 2, \dots$. Using the constraint $ -1 < \nu < 1$, 
it can be shown that for fixed values of
$N$, $g_1$ and $g_2$, there is at most one allowed value of $k$.

We encounter three kinds of singularities in the Hamiltonian
(\ref{e6}). They occur whenever,\\
(i) any two particles coincide, i.e. $x_i \rightarrow x_j$,\\
(ii) any particle is at the position of the image of any other particle,
i.e. $x_i \rightarrow - x_j$,\\
(iii) any particle is at the origin, i.e. $x_i \rightarrow 0$.\\
If we do not want the wavefunction to reflect singularities of these kinds,
we have to put the restriction $g_1, g_2 \geq 0$. For such cases, 
the new states
exists only in the $k=0$ sector and for $N=2$ and $3$. Compared to the
similar situation
in the case of $A_{N-1}$ Calogero model, $N=4$ is not allowed for the
present case. This may be attributed to the fact that $B_{N}$ Calogero
model is not translationally invariant.

In the more general case, following the discussion for the $A_{N-1}$
Calogero model, we allow the wavefunction to have singularities 
while maintaining the square-integrability of the angular part, 
which demands that $g_1, g_2 > -\frac{1}{2}$. 
We would now like to find the allowed values of $k$ when $N$ is kept fixed
but the couplings are allowed to vary. To this end, 
we first write $g_1$ and $g_2$ as,
\be
g_1 = - \frac{1}{2} + \epsilon_1^2, \ \ g_2 = - \frac{1}{2} + \epsilon_2^2,
\ee
\noindent so that $\epsilon_1$ and $\epsilon_2$ can take any nonzero 
real values. 
Using the constraint $ - 1 < \nu < 1 $, we find that
\be
\frac{1}{2} - \frac{2 k}{ N (N-1)} < \epsilon_1^2 + \frac{\epsilon_2^2}{N-1}
< \frac{1}{2} -\frac{2 k}{N(N-1)} + \frac{2}{N(N-1)}.
\label{new}
\ee
\noindent The upper bound in Eq. (\ref{new}) restricts $k$ as
\be
k < \frac{1}{4} N (N-1) + 1.
\ee
\noindent 
The allowed values of $k$ are given by $k=0, 1, 2, \dots,
K \equiv  \{\frac{1}{4} N (N-1)+ 1\}$.
When $k = K$, the l.h.s. of Eqn. (4.15) becomes negative if 
$\frac{1}{4} N (N-1)$ is not an integer.
In that case we replace the lower bound in Eqn. (4.15) by 0. 
It is evident from the above analysis that for a fixed value of $N$, 
new states exist for the range of the couplings given by
\be
0 < \epsilon_1^2 + \frac{\epsilon_2^2}{N-1} 
< \frac{1}{2} + \frac{2}{N(N-1)},
\ee
which defines an elliptical region in the $\epsilon_1 - \epsilon_2$ plane.
Since $g_1, g_2 > - \frac{1}{2}$, 
the lines given by $\epsilon_1 = 0$ and $\epsilon_2 = 0$ are excluded from
this elliptic region. 
For different allowed values of $k$, this region naturally separates into
disjoint elliptical shells defined by Eqn. (4.15). The outermost
shell corresponds to $k = 0$ while the innermost one corresponds to 
$k = K$. There are no new quantum states for the values of the couplings
corresponding to the boundary separating two consecutive shells. 

When $\nu = 0$,  the couplings satisfy the relation
\be
\epsilon_1^2 + \frac{\epsilon_2^2}{N-1}
= \frac{1}{2} + \frac{1 - 2 k}{N(N-1)} .
\ee
The allowed values of $k$ in this case are found to be
$k=0,1,\dots, \{(N^2 -  N + 2)/4 \}$. In general, a shell 
defined by Eqn. (4.15) contains an ellipse of the form (4.18) with the same
value of $k$. However, when $\frac{1}{4} N (N-1)$ is an integer, the value of
$\nu = 0$ cannot be achieved within the innermost shell corresponding to
$k = K$.

When $\nu \neq 0$, the bound state wavefunction of the $B_{N}$ model 
is given by 
\be
\psi = B  \prod_{i < j} (x_i^2 - x_j^2 )^{g_1} \prod_s
x_s^{g_2}
{\tilde {P}}_{2k}(x) r^{-\nu} H_\nu^{(1)}(i \sqrt{|E|}r),
\ee
where the bound state energy $E$ is given by Eqn. (3.13). ${\tilde {P}}$
appearing in Eqn. (4.19) is a solution of Eqn. (2.7) for the $B_N$ Calogero 
model. Similarly, the scattering state solutions are given by
\be
\psi = 
\prod_{i < j} (x_i^2 - x_j^2 )^{g_1} \prod_s x_s^{g_2}
{\tilde{P}}_{2k}(x) r^{-\nu} \left [ a(q) J_\nu (qr) - 
b(q) J_{- \nu} (qr) \right ],
\ee
where $\frac{a(q)}{b(q)}$ is given by Eqn. (3.9).
When $\nu = 0$, the bound state wavefunction is given by
\be
\psi = C \frac{i \pi}{2} 
\prod_{i < j} (x_i^2 - x_j^2 )^{g_1} \prod_s x_s^{g_2}
{\tilde{P}}_{2k}(x)  r^{-\nu}
H_0^{(1)} \left (i \sqrt{|E|} r \right ),
\ee
where the bound state energy $E$ are given in Eqn. (3.30).
Similarly, the scattering state solutions are given by 
\be
\psi = 
\prod_{i < j} (x_i^2 - x_j^2 )^{g_1} \prod_s x_s^{g_2}
{\tilde{P}}_{2k}(x) r^{-\nu} 
\left [ a(q)  J_0(qr) - b(q)  N_0(qr) \right ],
\ee
where ${a(q)\over b(q)}$ is given by Eqn. (3.27).

\noindent
{\bf 4.2.1 {\boldmath $D_N$} Calogero Model}

The $D_{N}$ Calogero model can be obtained from the $B_{N}$ model
by putting $g_2 = 0$, i.e. $\epsilon_2^2 = \frac{1}{2}$.
Using the expression for $\nu$ in (\ref{e8}) and the constraint
$ - 1 < \nu < 1$, we find,
\be
 - \frac{N+4 k}{2 N (N-1)} < g_1 <
- \frac{N+4 k}{2 N (N-1)} + \frac{2}{N(N-1)}.
\label{e9}
\ee
\noindent 
The constraint $ g_1 > -\frac{1}{2}$ restricts $k$ as 
\be
k \leq \frac{1}{4} ( N^2 - 2 N + 4 ).
\label{e10}
\ee
\noindent 
Thus, the quantum number $k$ can take any integral values,
$k=0, 1, 2, \dots, K \equiv \{\frac{1}{4} ( N^2 - 2 N + 4 )\}$.
It is clear that for a fixed value of $N$,
new states exist for the range of $g_1$ given by
\be
-\frac{1}{2} < g_1 < \frac{4-N}{2N(N-1)}.
\ee
Note, however, that the points given by $g_1 = -\frac{N + 4 k}{2N(N-1)}$
belonging to the range in Eqn. (4.25) are excluded. 
For fixed values of $g_1$ and $N$, the number of allowed values of $k$ is at 
most one. When $\nu \neq 0$, the bound and scattering state      
wavefunction in this case are obtained from Eqns. (4.19) and (4.20)         
respectively by putting $g_2 = 0$ in them.
When $\nu = 0$, $g_1 = - \frac{N + 4 k - 2}{2 N (N - 1)}$ with 
$k=0,1,\dots, \{(N^2 - N + 2)/4\}$. The bound and scattering states
wavefunctions in this case are obtained from Eqns. (4.21) and (4.22)
respectively by putting $g_2 = 0$ in them.

Finally, as in the case of $A_{N-1}$ Calogero model,
we have the remarkable results that for suitable ranges of couplings, 
the $B_{N}$ and $D_N$ Calogero models
admit self-adjoint extensions and consequently, new scattering and bound   
states for arbitrary $N$. These new states exist for certain non-zero angular 
momentum sector also. For fixed values of $N$ and the relevant couplings,  
$H_{A_{N-1}}$ admits a maximum of two bound states corresponding to at most
two possible values of $k$. On the other hand, under similar conditions,
$B_{N}$ and $D_N$ Calogero models admit only one bound state.
It may be noted that
the Hamiltonian of the $C_{N}$ Calogero model is identical to
$H_{B_{N}}$
except that the last term in Eqn. (4.11)is replaced by $ g_3 (g_3 -1 )
\sum_{i=1}^N (2x_i)^{-2}$ where $g_3$ is a constant. The analysis presented
above can be extended to the $C_{N}$ Calogero model as well.

\noindent
{\bf 4.3 Calogero-Marchioro Model : A {\boldmath $D$} Dimensional Example}

All the models that have been discussed so far are one dimensional. 
We now give an
example of a $D$ dimensional many-particle system, known as the
Calogero-Marchioro model in the literature \cite{cm,pkg}. The Hamiltonian
for this model is given by
\be
H_{CM} = - \sum_{i=1}^N \grad_i^2 + g (g + D -2) \sum_{i \neq j}
\frac{1}{{\bf r}_{ij}^{2}} + g^2 \sum_{i \neq j \neq k}
\frac{ {\bf r}_{ij} \cdot {\bf r}_{ik} }
{{\bf r}_{ij}^{2} {\bf r}_{ik}^{2}}, 
\label{e12}
\ee
\noindent where ${\bf r}_{ij} = {\bf r}_i - {\bf r}_j$. The Hamiltonian
in Eqn. (4.26) can be obtained from D-dimensional analogue of 
Eqn. (\ref{p1}) by choosing
\be
G = \prod_{i \neq j} {\bf r}_{ij}^g.
\label{e13}
\ee
\noindent 
$G$ is a homogeneous function of degree $d= g N (N-1)/2$. $H_{CM}$ 
contains both two-body and three-body interaction terms. For $D=1$, the
three-body interaction term vanishes identically and $H_{CM}$ reduces to the
the $H_{A_{N-1}}$. For $D \geq 2$, although infinitely many exact
solutions corresponding to the radial excitations are known, the model is
not exactly solvable. In particular, Eq. (\ref{p7}) determining solutions
for $P_k$ is not known for $ k > 0$. We will henceforth restrict our attention
to the $k=0$
sector of the model. It may be mentioned here that the Calogero-Marchioro
model in $D=2$ is related to several systems of physical interest, e.g.
quantum Hall effect, quantum dots, random matrix theory etc. \cite{pkg}.
Moreover, $H_{CM}$ in $D=2$ can be embedded into a Hamiltonian with 
extended ${\cal{N}} = 2$ superconformal symmetry \cite{pkg}.

The Hamiltonian in Eqn. (4.26) 
is translationally invariant and the parameter $\nu$ is
determined from Eqn. (2.17) as 
\be
\nu = \frac{1}{2} (N-1) D - 1 + \frac{g}{2} N ( N-1).
\label{e14}
\ee
\noindent The many-body interactions are singular at the coincident positions
of any two particles. If we do not want the wavefunction to have
singularities of this kind, we have to impose the condition $g > 0$, which
requires that $0 < \nu < 1$. Under this condition, for fixed values of $N$
and $D$, $H_{CM}$ admits self-adjoint extensions when
\be
1 + \frac{2}{D} \leq N < 1 + \frac{4}{D}.
\label{e15}
\ee
\noindent 
First note that the result of the $A_{N-1}$ Calogero model is reproduced 
for $D=1$. For $D=2$, only $N=2$ is allowed and
for $ D > 3$, there are no valid solution.

In the more general case, we allow the wavefunctions to have singularities
while maintaining the square-integrability of the angular part coming from
$G$. This puts the restriction that $ g > -\frac{1}{2}$. In this case $\nu$
satisfies the constraint $-1 < \nu < 1$ which restricts $g$ as
\be
- \frac{D}{N} < g < - \frac{D}{N} + \frac{4}{N (N-1)}.
\label{e16}
\ee
\noindent 
Let us first consider the case $\nu \neq 0$. 
The condition $g > - \frac{1}{2} $ is satisfied  for
\be
D < \frac{N}{2} + \frac{4}{N-1}.
\label{e17}
\ee
\noindent 
The self-adjoint extension and consequently, the existence of
scattering and bound states are admissible, provided Eqns. (\ref{e16})
and (\ref{e17}) are satisfied simultaneously. For $ D > 3 $, there exists
a lower bound on $N$ in order to have the self-adjoint extension. However,
interestingly enough, for the physically important cases of $D=1,2$ and $3$,
the self-adjoint extensions are admissible for any arbitrary $N$. 

When $\nu = 0$, the coupling $g$ takes the value given by
\be
g = \frac{2}{N(N-1)} - \frac{D}{N}.
\ee
The constraint $g > -\frac{1}{2}$ leads to the condition
\be
D < \frac{2}{N-1} + \frac{N}{2}.
\ee
Note that Eqn. (4.33) is satisfied for arbitrary $N$ when $D = 1,2$.
For $D=3$, it is valid for $N \geq 6$. We end this discussion by noting that
the wavefunctions and the spectrum for this model
can be written explicitly as before, which we do not discuss here in detail.

\sxn{Conclusion}

	In this paper we have discussed the quantization of $N$ particle
systems with classically scale invariant long range interactions.
The effective Hamiltonian for these
systems in the ``radial'' direction contains an inverse square term whose
coefficient $(\nu^2 - \frac{1}{4})$ depends on the particle number $N$, 
the couplings and the ``generalized angular momentum'' $k$. It has been found 
that when $ -1 < \nu < 1$, the Hamiltonians for this entire class of systems 
admit a one parameter family of self-adjoint extensions labeled 
by $e^{iz}$ where $z \in R$ (mod $2 \pi$). 
The parameter $z$ classifies all possible boundary
conditions for which the Hamiltonian is self-adjoint. Moreover, the spectrum
of the Hamiltonian depends explicitly on $z$. Each choice of the 
parameter $z$ thus gives rise to an inequivalent quantization of the system.

We have illustrated the general approach through several examples
which have been studied in detail. The examples in one dimension 
include $A_{N-1}$ and $B_{N}$ Calogero models without the 
confining term. The $D_{N}$ Calogero model has been studied
as a special case of the $B_{N}$ case. In these cases we have found that 
a new class of bound and scattering states appear for arbitrary values of
the particle number $N$ and within certain ranges of the coupling constants. 
Moreover, these new states appear not only in the $k=0$ sector, but for
higher values of the quantum number $k$ as well. To our knowledge, this is
the first demonstration of the existence of self-adjoint extensions in the
excited sectors of a system. This result may be attributed to the highly
correlated nature of the many-body interaction. 
It may be mentioned that there is an important difference 
between the spectrum of $H_{A_{N-1}}$ and $H_{B_{N}}$. 
For fixed values of $N$ and the relevant couplings,    
$H_{A_{N-1}}$ admits a maximum of two bound states corresponding to   
two allowed values of $k$. On the other hand, under similar conditions,     
$B_{N}$ Calogero models admit only one bound state. This can be attributed
to the fact that $H_{A_{N-1}}$ is translationally invariant, while $H_{B_{N}}$
is not.

As an example of our general discussion in dimension higher than one, we have
analyzed the Calogero-Marchioro model in $D$ dimensions.
In this case, the solution of the angular equations is known only for $k=0$ 
and we have presented the analysis only for this sector. For the physically
important cases of $D = 1,2$ and $3$, the new
states are found to exist for arbitrary number of particles $N$ and within 
certain range of the coupling constant.

Although the class of systems considered here is classically
scale invariant, we have found
that in presence of the self-adjoint extension, the
systems may admit bound states. Moreover, the associated $S$ matrix and the
phase shifts are found to depend explicitly on the momentum. This happens as 
the classical scale invariance is broken due to quantization, which is
manifest by the fact
that the scaling operator does not leave the domain of self-adjointness of
the Hamiltonian invariant. The analysis here thus provides further examples of
quantum mechanical scaling anomaly. We have also shown that scale invariance
at the quantum level can be implemented for special choices of the parameter
$z$. It may also be mentioned in this context that the effective Hamiltonian 
$\tilde{H}$ in Eqn. (2.11) can be shown to be related to the Virasoro algebra 
\cite{kac} with central charge $c=1$ \cite{kumar1,kumar2}. The systems
studied here would also be related to the Virasoro algebra.

We have restricted our attention to the case where the coefficient
of the inverse square potential is not too strongly negative in order to
avoid the ``fall to the center'' \cite{land}. The strong coupling region of
the analogous 2-body problem has been analyzed in the literature
using renormalization group techniques \cite{ksg}. It would be interesting to 
consider the problem of renormalization  in the presence of the self-adjoint
extension. The N-particle rational Calogero models are also related to 
black holes \cite{gib} and
Yang-Mills theories \cite{poly,lang}. It is plausible that the quantum states 
found here would have analogues in those cases as well. 
It would also be interesting to investigate the Calogero models associated
with the exceptional Lie groups \cite{pr}. Finally we would like
to mention that an analysis similar to the one presented in this paper can
be done in the case of rational Calogero models in the presence of the
confining term as well \cite{fut}.

\vskip 1cm
\noindent
{\bf Acknowledgements}\\

 The work of PKG is supported(DO No. SR/FTP/PS-06/2001) by
the SERC, DST, Govt. of India, under the Fast Track Scheme for Young
Scientists:2001-2002.

\bibliographystyle{unsrt}

\end{document}